\documentclass[twocolumn,showpacs,preprintnumbers,amsmath,amssymb]{revtex4}
\usepackage{dcolumn}
\usepackage{bm}
\usepackage[dvipdf]{graphicx}
\DeclareGraphicsExtensions{.jpg,.pdf,.mps,.png,.eps,.ps,.EPS}
\begin{document}
\newcommand{\avg}[1]{\langle{#1}\rangle}
\newcommand{\Avg}[1]{\left\langle{#1}\right\rangle}
\def\be{\begin{equation}}
\def\ee{\end{equation}}
\def\bea{\begin{eqnarray}}
\def\eea{\end{eqnarray}}
\title{A statistical mechanics approach for scale-free networks and finite-scale networks }
\author{Ginestra Bianconi }
\affiliation{The Abdus Salam International Center for Theoretical Physics, Strada Costiera 11, 34014 Trieste, Italy } 
\begin{abstract}
We present a statistical mechanics approach for the description of complex networks. We first define an energy and an entropy associated to a degree distribution which have a geometrical interpretation. Next we evaluate the distribution which extremize the free energy of the network. We find two important limiting cases: a scale-free degree distribution and a finite-scale degree distribution. 
The size of the space of allowed simple networks given these distribution is evaluated in the large network limit. Results are compared with simulations of algorithms generating these networks.
\end{abstract}
\maketitle
{\bf Universality features of the structure of complex networks are very good candidate for a statistical mechanics treatment. While  non-equilibrium aspects of these networks was first elucidated,  later attention has been addressed to equilibrium properties of these graphs. In this paper we reveal some new property of finite-scale and scale-free networks. We characterize their different degree distribution as equilibrium degree distribution of a statistical mechanics problem written in terms of an energy and an entropy having a graphical interpretation. We observe that scale-free degree distribution  for uncorrelated networks and in the large size limit correspond to structure in which the space of distinguishable simple graphs  is strongly suppressed respect to the finite-scale distribution case.    }

Complex networks describe the intertwin relations between the elements of a variety of complex systems  as different as  the Internet and the biological networks of the cell \cite{RMP,Doro,Vespi}.
To understand the interplay between their functions and their structure it is natural to  study how far from optimal performance are these networks. Consequently optimal networks have been defined respect  to a specific function \cite{Maritan,Flammini,Newman1},  or {\em ii)} to their dynamics \cite{Vega,Miguel,Motter}, or {\em iii)} to  some  topological robustness features \cite{Two_p}.
Nevertheless to understand the universal features of these networks which are found in many networks regardless of the specific dynamics which takes place of them, statistical mechanics approaches are more appropriate.
A major universal property of complex networks is their  degree distribution $P(k)$. In fact both scale-free networks $P(k)\sim k^{-\gamma}$ \cite{BA} with diverging second moment of the degree distribution $\avg{k^2}$ and finite-scale networks with finite second moment $\avg{k^2}$ are widely encountered in many real networks.
 These two general network distributions have very strong impact also on the dynamics defined on these graphs as it has been widely discussed in the literature  \cite{effe1,effe2}.

Different statistical mechanics approaches have been used to characterize complex (scale-free and finite-scale) networks.
In particular the attention has been first addressed to the out-of-equilibrium dynamics which by the means of preferential attachment \cite{BA} generates scale-free degree distributions. This dynamics is justified by direct measurements for example in the case of the  the World-Wide-Web \cite{WWW}.
Secondly,  different  approaches have been taken to assess if the scale-free degree distribution has to be considered as the equilibrium distribution of some statistical mechanics problem \cite{Palla,Burda,static,hv,Berg,Burda3,Dorogovtsev,Palla,Newmanb,Ohkubo}.  
If we restrict ourself to uncorrelated networks  we see that the approaches can be classified into three major classes:
the ones which see the emergence of a scale-free network structure at the phase transition   \cite{Burda,Palla};
the ones which assume a sort of preferential attachment in the dynamics \cite{Ohkubo,Dorogovtsev}
the ones which consider some hidden variables associated to each node of the network s\cite{hv,static,Newmanb}.
Models in the first two classes have strong relations.

In particular in \cite{Palla}  an energy of the type $E=-\sum_i k_i\log(k_i)$ was considered. The Montecarlo probability for a single move at $T=1$ involve a linear preferential attachment mechanism with nodes of higher degree attracting links with higher probability. As a function of the temperature of the graph the model  has two phase transitions. At the boundary between a sparse phase and a connected phase at a critical temperature $T_c\sim 1$   a scale-free network  with exponent $\gamma=3$ is found .
In \cite{Burda} the authors define a minifield theory whose Feynman diagrams are the networks of the ensemble. The networks of these ensemble  map to a urn or "ball in the box" model \cite{Burda2} in which the ``boxes'' map to the nodes, the ``balls'' map to the edges of the graph and the probability of having $k$ edges at node $i$ is given by $p(k)\propto k^{-\gamma+1}$.The scale-free network ensemble is recovered  in the case of trees or networks with finite number of cycles. The same approach is extend further to graphs with cycles in \cite{Burda,Burda3}, in which the authors show that the mapping to the "ball in the box" model regards a distribution  $p(k)\propto k^{-\gamma}.$ In the canonical ensemble of Ref. \cite{Dorogovtsev}, the authors describe an equilibrium network in which there is preferential attachment of nodes of high degree. The model again map to the urn model \cite{Burda2} with the probability $p(k)\sim k^{-\gamma}$.
Finally another urn ("ball in the box" model ) is presented in \cite{Ohkubo} in which the same mapping between the network and the ``ball/boxes'' is made but the probability $p(k)=(k!)^{\beta}$ is chosen. This choice of the $p(k)$ probability involve some preferential attachment of the dynamics when the heat-bath rule \cite{Godreche} is considered. A finite scale network if recovered when $\beta=1$ for each node of the graph while a scale-free degree distribution  with exponent $\gamma=2$ plus logarithmic corrections is obtained  in the case of uniform distribution of effective $\beta_i$ on each node $i$ of the network.

In the following paper we will present a statistical mechanics treatment of complex networks in which we consider as thermodynamic quantities two terms having a precise graphical interpretation. The ``energy'' associated with the degree distribution will the  logarithm of the number of allowed equivalent simple networks is possible to draw given a degree sequence. The ``entropy'' associated with a   degree distribution will be the logarithm of the number of ways in which we can distribute  $L$ edges among the $N$ nodes of the graph following the distribution $\{N_k\}$.  
From this statistical mechanics approach we observe the  emergence
either of  scale-free networks or finite-scale networks without  been
close  to a phase transition in the network. The models is very closed
related to the urn ("balls in the box") models but with urn ("boxes") mapping to the degree of the nodes and not to the nodes themselves.
On the same time  the dynamical rules which give rise to the network directly derives from the choice of the energy function we make. The appearance of scale-free degree distribution in the opposite limit of some finite-scale degree distributions can be put in relation with the  findings of Ref.  \cite{Bose,Fermi,Mixed} where in the contest of growing networks  it was shown that the growing Cayley tree networks are described by opposite equations with respect to the growing scale-free networks. Nevertheless, here the network is not growing and the nodes are all equivalent since there is no fitness associated  to the nodes. 
Although the dynamics mechanism giving rise to these distributions do not appear to be related to real phenomena occurring in evolving networks (as for example the preferential attachment mechanism) this approach has the advantage to characterize uncorrelated scale-free networks and finite scale networks on the base of the multiplicity of simple graphs is it possible to draw given a degree distribution. 
In fact  we obtain  that scale-free degree distributions characterize  networks in which the energy is much higher  than for finite-scale networks. In the paper we will show that this implies for uncorrelated simple graphs  that  the number of allowed distinguishable simple graphs given a certain degree distribution is minimized for scale-free networks.  This result indicates that in the swapping algorithm for randomizing a network \cite{Sneppen} the number of different graphs one can  sample is actually reduced for scale free networks with low exponent $\gamma$.

In order to define our partition function we consider a network with $N$ nodes and $2L$ edges.
We distribute  the edges $e=1,\dots 2L$, through  the $N$ nodes of the network and we assign to each edge a variable $s_{e}=1,\dots,K$  indicating the degree of the assigned node. The degrees $k$ takes the values $k=1,\dots,K$ with maximal allowed degree $K$. The degree distribution $\{N_k\}$ of a single assignment $\{s_{e}\}$ is such that the total number of edges departing from nodes of degree $k$ can be expressed as $k N_k=\sum_{e=1}^{2L} \delta_{k,s_{e}}$.
We would like to study which are the equilibrium properties of the distribution of $\{s_{e}\}$ that maximize or minimize an energy functional   $E(\{s_e\})$ .
Therefore, considering a problem close to the original Backgammon model \cite{Ritort}, we define the partition function 
\be
Z=\frac{1}{(2L!)}\sum_{\{s_{e}\}}{}^{'} e^{-\frac{1}{z} E(\{s_e\})}
\label{Z0.eq}
\ee
where the normalization factor $(2L)!$ is chosen for convenience and the sum is extended over all distributions $\{s_{e}\}$ with $e=1,\dots, 2L$ such that satisfy the conditions
\bea
kN_k=\sum_{e} \delta_{s_{e},k}\nonumber \\
\sum_k N_k=N.
\eea
In the following we will  consider an the energy  $E(\{s_e\})=E (\{N_k\})$ associated to the degree distribution of the network given by  the logarithm of the  number ${\cal N}_{G}$ of indistinguishable simple networks it is possible to draw given the degree sequence, i.e. 
\be
E (\{N_k\})=\log({\cal N}_G),  
\ee
(where simple networks are  networks without tadpoles and double links).
The number  ${\cal N}_G$ can be expressed as 
\be
{\cal N}_G=e^{E (\{N_k\})}={\prod_k k!^{N_k}}.
\label{S_g.eq}
\ee  
In fact, for every simple graph associated with the given distribution every  permutation of the edges departing from each node generates the same set of links. These permutations are given by $\prod_k k!^{N_k}$, and consequently we derive Eq. $(\ref{S_g.eq})$.
 For $z>0$  low energy configurations are weighted more  while  for $z<0$ high energy configurations are more weighted in the  partition function $Z$ defined in Eq. \ref{Z0.eq}. Consequently we expect that in the system for $z<0$ we should bound the energy per particle and therefore the maximal connectivity to a finite value $K$.
Since the energy associated to the distribution $\{s_{p}\}$ only depends on the degree distribution $\{N_k\}$ we can  write the partition function as a sum over the  distributions $\{N_k\}$
 with given energy $E (\{N_k\})$ and entropy $S (\{N_k\})$, i.e.
\be
Z=\frac{1}{(2L)!}\sum_{ \{N_k \} }{}^{'}  e^{-\frac{1}{z} E (\{N_k\})+S(\{ N_k\})}.
\label{Z.eq}
\ee
The  entropic term $S (\{N_k\})$  is equal to   the logarithm of the number of ways ${\cal N}_{\{N_k\}}$ in which we can distribute  $2L$ edges into any degree sequence $\{k_1,\dots, k_N\}$ of distribution $\{N_k\}$, or equivalently the number of sequences $\{s_{e}\}$ which correspond to a degree distribution $\{N_k\}$ and is  given by
\be
e^{S (\{N_k\})}={\cal N}_{\{N_k\}}=\frac{(2L)!}{ \prod_k (k N_k)! }.
\ee
For $z>0$ the equilibrium degree distribution   $\{N_k\}$ will minimize the free energy of the network $F(\{N_k\})= E (\{N_k\}) -z S(\{ N_k\})$. For $z<0$ the equilibrium distribution will maximize the free energy $E (\{N_k\}) -z S(\{ N_k\})$. 
The role of the parameter  $z$ is to measure  a tradeoff between the 'energetic' and the 'entropic' term in the definition of the free energy, as well as the temperature  $T$ in classical statistical mechanics.  Respect to the other statistical mechanics problems described in the introduction, this  model can be cast into  a "ball in the box" model with ball mapping to edges and ``boxes'' mapping to connectivity values and a weight $p(k)=(k!)^{-\beta/k}$ associated to each edge of the graph linked to a node of degree $k$  with $p(k)\propto k^{-\beta}$ for large values of $k$.  
In equation $(\ref{Z.eq})$ the sum $\sum'$ over the $\{N_k\}$ distributions is extended only to  $\{N_k\}$ for which  the total number of nodes $N$ and the total number of links $L$ in the network is fixed, i.e.
\bea
\sum_k N_k=N \nonumber \\
\sum_k k N_k=2 L.
\label{conditions}
\eea
To enforce these conditions we introduce in $(\ref{Z.eq})$ the delta functions in the integral form obtaining the expression
\bea
Z&=&\frac{1}{(2L)!}\int \frac{d\lambda}{2\pi} \int \frac{d\nu}{2\pi} \sum_{ \{N_k \} } \exp\left[ -\frac{1}{z} E (\{N_k\})+S(\{N_k\})\right.\nonumber \\
& &\left.-i\lambda
(2L-\sum_k k N_k)-i\nu (N-\sum_k N_k)\right].\nonumber \\
&=&\int \frac{d\lambda}{2\pi} \int \frac{d\nu}{2\pi} \exp\left[-i\lambda 2L -i\nu N+ \sum_k \log G_k(\lambda,\nu)\right]=\nonumber \\
&=&\int \frac{d\lambda}{2\pi} \int \frac{d\nu}{2\pi} \exp[Nf(\lambda, \nu)]
\label{Z2}
\eea
where
\be
G_k(\lambda, \nu)=\sum_{N_k} \frac{1}{(k N_k)!} \exp\left\{k N_k\left[i\lambda +i \frac{\nu}{k}-\frac{1}{z k}\log(k!) \right]\right\}.
\ee
Assuming that the sum over all $N_k$ can be approximated by the sum over all $L_k=k N_k=1,2,\dots \infty$ we get $\log G_k(\lambda, \nu)=\exp\left[{i\lambda +i\nu/k -\frac{1}{kz}\log(k!)}\right]$ and
\be
f(\lambda, \nu)=-i\avg{k}\lambda-i\nu+\frac{1}{N}\sum_k e^{i\lambda+i\nu/k -\frac{1}{zk}\log(k!)},
\ee
where $<k>=2L/N$ indicates the average degree of the network.
By evaluating $(\ref{Z2})$ at the saddle point, deriving the argument of the exponential respect to $\lambda$ and $\nu$, we obtain 
\bea
\avg{k}=\frac{1}{N}\sum_k  e^{i\lambda+i\nu/k -\frac{1}{zk}\log(k!)} \nonumber \\
1=\frac{1}{N}\sum_k  \frac{1}{k} e^{i\lambda+i\nu/k -\frac{1}{zk}\log(k!)}
\label{dOmega}
\eea
and the marginal probability that  $L_k=k n=\ell$ is given by
\be
P(L_k=\ell=nk)=\frac{1}{\ell!}e^{-\ell/k\log(k!)}\frac{Z_k(L,\ell,N)}{Z(L)},
\label{marginal}
\ee
with
\be
Z_k(L,\ell,N)=\int \frac{d\lambda}{2\pi} \int \frac{d\nu}{2\pi} \exp[Nf_k(\lambda, \nu,\ell)]
\ee
and $f_k(\lambda,\nu,\ell)=i(\avg{k}-\ell/N)\lambda-i\nu(1-\ell/(kN))+\frac{1}{N}\sum_{s\neq k} \exp[{i\lambda+i\nu/s -\frac{1}{zs}\log(s!)}].$
If we develop $(\ref{marginal})$  for  $\ell \ll L $ and we use the Stirling approximation for factorials,  we get that each variable $L_k$ is a Poisson variable with mean $<L_k>$ satisfying
\be
\frac{<L_k>}{k}=<N_k>=k^{-\frac{1}{z}-1}e^{\lambda+\frac{1}{z}} e^{\nu/k}.
\label{Nk}
\ee
\begin{figure}
\begin{center}
\includegraphics[width=65mm, height=55mm]{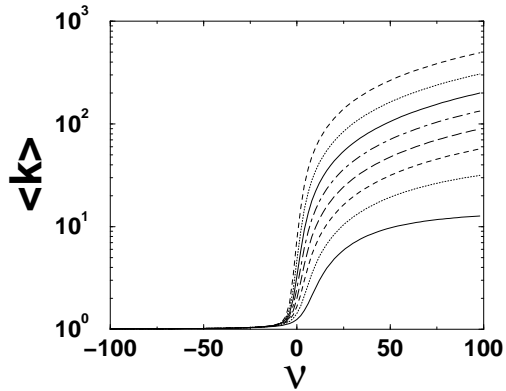}
\caption{Average degree $\avg{k}$ of the  distribution $<N_k>$  as a function of $\nu$ for  networks with different value of $z$, starting with $z=0.2$ (solid line at the bottom), and from bottom to top the curves with $z=0.3,0.4,0.5,0.6,0.7,0.8$ are drawn  until the curve with $z=0.9$ (dashed line at the top).}
\label{kav.fig}
\end{center} 
\end{figure}
If we restrict ourself to the networks with finite average degree in the thermodynamic limit, the allowed values of $z$ are $z\in(-1,1)$.
From the expression $(\ref{Nk})$ for $<N_k>$, if $z\in(0,1)$ the equilibrium degree distribution is scale-free  with a power-law tail characterized by the exponent $\gamma=\frac{1}{z}+1$.  The parameter $\nu \neq 0$ modulates  the average degree of the graph constituting for $\nu>0$ an effective lower cutoff of the distribution wherever  the upper  cutoff $K$ of the degrees is  the natural cutoff of the  distribution $(\ref{Nk})$.
A different scenario arises  if $z<0$, when the equilibrium network, in average  $(\ref{Nk})$   has a  power-law degree distribution increasing with the degree $k$. In this case the Lyapunov functions $\nu $ and $\lambda$ cannot fix the average degree unless one introduces by hand an upper cutoff  $K$ in the degree of the nodes of the order of magnitude of the average degree $\avg{k}$. We note here that also for $z\in (0,1)$ it could be convenient to set by hand a structural cutoff $K\sim N^{1/2}$ for $z>1/2$ in order to obtain an uncorrelated network. 

 The  values of the Lagrangian multipliers $\lambda$ and $\nu$ and of the 
upper cutoff $K$ are fixed  in the case  $z\in (0,1)$ by the conditions:
\bea
\nu^{-\frac{1}{z}} e^{\lambda+\frac{1}{z}}\left[\Gamma(\frac{1}{z},\frac{\nu}{K})-\Gamma(\frac{1}{z},\nu) \right] &=&N\nonumber \\\nu \left[\Gamma\left(\frac{1}{z}-1,\frac{\nu}{K}\right)-\Gamma\left(\frac{1}{z}-1,{\nu}\right)\right] &= &\nonumber \\=\avg{k}\left[{\Gamma\left(\frac{1}{z},\frac{\nu}{K}\right)-\Gamma\left(\frac{1}{z},\nu\right)}\right]&&\nonumber \\
\nu^{-\frac{1}{z}} e^{\lambda+\frac{1}{z}}\left[\Gamma(\frac{1}{z})-\Gamma(\frac{1}{z},\frac{\nu}{K}) \right] &=&1,
\label{Pl.eq}
\eea
where $\Gamma(a,b)$ indicates the incomplete Gamma function and $\avg{k}=2L/N$ the average connectivity of the network. 
This system of equations is solvable provided $\avg{k}>1$, as can be seen from  Figure $\ref{kav.fig}$  where we plot the average connectivity of a network as a function of $\nu$ for different values of $z$.
\begin{figure}
\includegraphics[width=75mm, height=115mm]{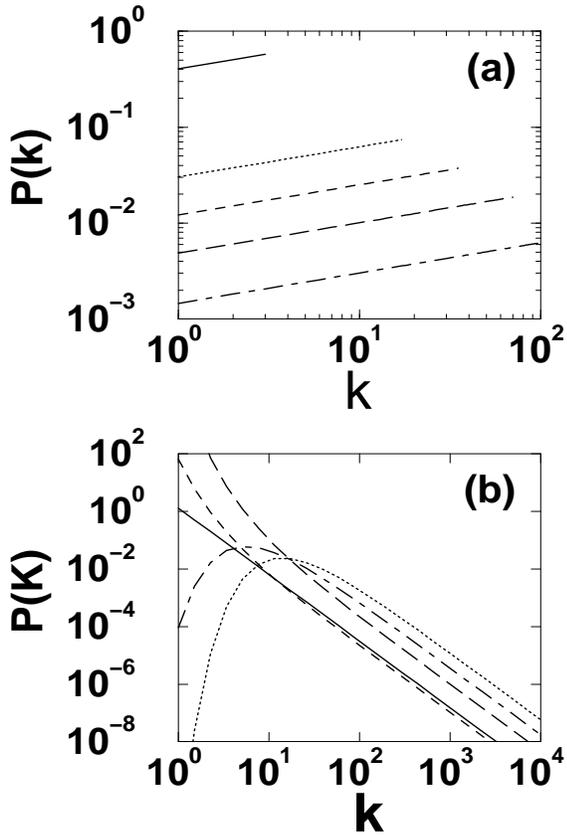}
\caption{The  distribution $<N_k>$ for $z=-0.75$ (panel {\bf(a)}) and $z=0.75$  (panel {\bf  (b)}) as a function of the average degree of the network. Data are shown for $\avg{k}=2,10,20,40,100$(dotted solid line) for $z=-0.75$ and for $\avg{k}=1.3,2,$, $\avg{k}=4$ (solid line), $\avg{k}=40,100$ for $z=0.75$.}
\label{stat_new.fig} 
\end{figure}

On the other hand, in the case  $z\in (-1,0)$ we have to fix the upper cutoff $K$ for the system to be stable and not crumpled.
If we take $K=(1+z) \avg{k}$
the parameters $\nu,\lambda$ satisfy the following conditions,
\bea
\nu&=&0\nonumber \\
e^{\lambda+\frac{1}{z}} K^{\frac{1}{z}} z &=&N.\nonumber \\
\label{Fs.eq}
\eea 
The two types of distributions described by $(\ref{Nk})$ with $z>0$ and the parameter of the distribution satisfying Eqs. $(\ref{Pl.eq})$, or with $z<0$ and the parameter of the distribution satisfying Eqs. $(\ref{Fs.eq})$, are significantly different.
In fact, for  $z>0$   there are highly connected nodes in the network and  the degree distribution for large degrees $k$ decays as a power-law with an  exponent $\gamma$ fixed by the value of $z$.
On the contrary,  for  $z<0$  the predicted distribution $(\ref{Nk})$  has a finite-scale and there are no highly connected nodes. Furthermore the topology of the network is very different from the one of the power-law case since the most connected nodes are also more abundant  than less connected ones.
In Figure $\ref{stat_new.fig}$ we show the  distributions $<N_k> $which solve these equations  for $z$ positive and negative at different values of the average connectivity $\avg{k}$ taking $K=(1+z)\avg{k}$ for $z<0$.

\begin{figure}
\includegraphics[width=65mm, height=55mm]{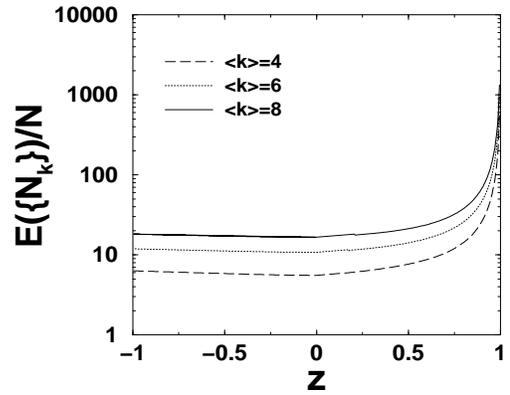}
\caption{Entropy of the network as a function of the parameter $z$,  in the thermodynamic limit $N\rightarrow \infty$ }
\label{S_g.fig} 
\end{figure}

Given the  distributions $(\ref{Nk})$  we can calculate the energy of the network as a function of $z$ at fixed average connectivity $\avg{k}$, always fixing the upper cutoff to $K=(1+z)\avg{k}$.
In Figure $\ref{S_g.fig}$ we present the energy of the network such that $N_k=<N_k>$ as a function of $z$ for different average connectivities.

The energy has a minimum in the limit  $z\rightarrow 0$ when the equilibrium degree distribution is such that  $<N_k>$ in infinitely peaked around the average connectivity. 
On the contrary, in the limit  $z\rightarrow 1$ where  the degree distribution has a  power-law exponent $\gamma\rightarrow 2$ the energy $E (\{N_k\})$  is at the maximum.
We note here that the energy $E (\{N_k\})$ of complex networks  calculated on  the equilibrium degree distribution $(\ref{Nk})$, is an extensive quantity in  both cases $z>0$ and $z<0$.

In the case $z<0$ and in the case $z>0$  where we introduce by hand a structural cutoff $K\sim N^{1/2}$ we can assume that the networks described in this paper  are randomly wired. In this case we can evaluate the number ${\cal N}_{SG}$ of distinguishable simple graphs it is possible to construct  given the degree distribution $(\ref{Nk})$.
This number is approximated by
\be
{\cal N}_{SG}\propto\frac{(2L)!!e^{-\frac{1}{2}\left(\frac{<k^2>}{<k>}\right)^2}}{e^{E(\{N_k\})}}
\label{N_SG}
\ee
In fact the total number of wiring it  is possible to draw given $2L$ edges is given by $(2L)!!$. This number include all type of possible wiring of the edges including the ones which give rise to graphs which are not simple.
Assuming that the graph is randomly wired, i.e. that the probability that a node with $k_i$ edges connect to a node with $k_j$ edges is a Poisson variable with average  $k_i k_j/(<k>N)$ the probability $\Pi$ that the graph is simple is equal to  \cite{loops_lungo}
\be
\Pi=\prod_{i, j} \left(1+\frac{k_i k_j}{<k>N} \right)e^{-{k_i k_j}{<k>N}}\sim e^{-\frac{1}{2} \left(\frac{<k^2>}{<k>}\right)^2}.
\ee
Finally in the expression $(\ref{N_SG})$ for ${\cal N}_{SG}$ there is an additional terms which takes into account the equivalent wiring of the edges which is given by $e^{-E(\{N_k\})}$.
The term $\frac{<k^2>}{<k>}$ for scale-free graphs with cutoff $K\propto N^{1/2}$ is subleading respect to the energetic term $E{(\{N_k\})}$ which dominates for large network sizes $N$.
Consequently the number $\cal N_{SG}$ of distinguishable simple graphs given a degree sequence is maximal for the graph with $z=0$ and minimal  for the scale-free graph with $\gamma \rightarrow 2$.

\begin{figure}
\begin{center}
\includegraphics[width=75mm, height=45mm]{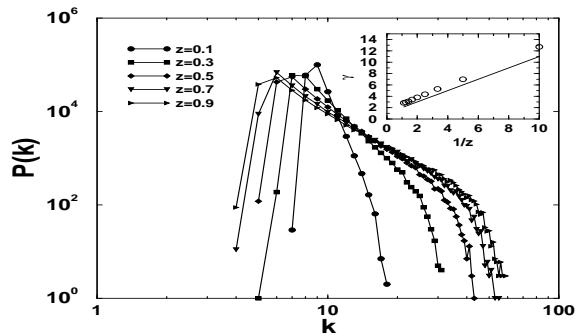}
\caption{MonteCarlo results for networks of $N=1000$ nodes and $L=4000$ links for different value of $z$. The Inset report the power-law exponent $\gamma$ of the distribution as a function of $z$ (empty circles) and the predicted  behavior (solid lines).}
\label{MonteCarlo.fig} 
\end{center}
\end{figure}
\begin{figure}
\includegraphics[width=65mm, height=55mm]{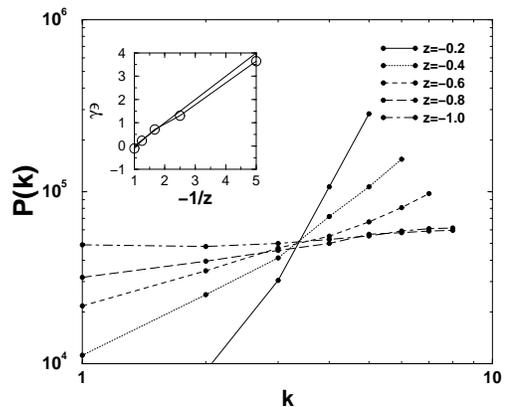}
\caption{Degree distribution of the network evolving following algorithm II with different values of $zeta$. The networks  have $N=500$ nodes $L=1000$ links have been evolved in $5000 L$ timesteps for $10$ runs and have a maximal cutoff $K$ taken to be $K=[\avg{k}(1+z)]$. }
\label{fspk.fig} 
\end{figure}
\section{Algorithms}

From the derivation of our model it is evident the connection  with ``ball in the  box'' problems \cite{Ritort,Burda2,Ohkubo}. 
In our approach the "boxes" map to   the degree of the nodes and the ``balls'' map  to the edges of the graph. This makes a crucial difference respect with the models  \cite{Burda,Ohkubo},  in which  the "boxes" map to the nodes of the graph. 
The model, as well as a urn model can be simulated using a heat-bath rule \cite{Godreche} which then suggest the following  algorithm: \\
\\
-Algorithm I-\\
\begin{itemize}
\item Choose   randomly a "ball" randomly, i.e. choose a link $(i,j)$;
\item Choose  a ``box'' randomly, i.e. chose a node $j'$ with a random value of its degree $k_{j'}$. This choice is implemented by  choosing a node $j'$ with probability proportional to $1/N_{k(j')}$.  
\item Swap link $(i,j)$ with link $(i,j')$ with the heat bath rule algorithm, i.e. with probability $\Pi=(k_{j'}+1)^{-1/z}$ 
\end{itemize}
From the shape of the linking probability $\Pi$ we infer that this algorithm,  include some sort of inverse preferential attachment  since rewiring to nodes with lower degree are more frequents.
In   Figure $\ref{MonteCarlo.fig}$ we report the resulting degree distributions as a function of $z$. The resulting distribution differ from the theoretical prediction for the presence of sharp lower cutoff. Nevertheless  the distribution for large $k$  decays with  power-law exponents close to the expectation (see Inset Figure \ref{MonteCarlo.fig}) with the difference depending on  finite size effects.
We checked that  the upper cutoff $K$ scale with the network size as $K \sim N^{\alpha}$ and $\alpha\propto z$.   

For the case $z<0$ the algorithm follow a rule similar to the one used in the case $z>0$ with the further introduction  of the ad hoc upper cutoff $K$ which is needed for the stability of the algorithm.
Consequently the algorithm is as following:\\
\\
-Algorithm II-\\
\begin{itemize}
\item Choose   randomly a link $(i,j)$;
\item Choose  node $j'$ with a random value of its degree $k_{j'}$, i.e. choose a node $j'$ with probability proportional to $1/N_{k(j')}$.  
\item Swap link $(i,j)$ with link $(i,j')$ only if $k_{j'}<K$ and in this case  with probability $\Pi$ 
\end{itemize}
with
\be
\Pi=\frac{(k_{j'}+1)^{-1/z}}{\sum_{r=1}^N (k_{r}+1)^{-1/z}}.
\ee
The results of the simulations following algorithm II are reported in Fig. $\ref{fspk.fig}$. The algorithm requires a very long equilibration time respect to the case with $z>0$. Consequently we consider relatively small networks sizes up to $500$ nodes. 
In the Inset we report the value the best value of the power-law fit to data together with the theoretical expectations.


In conclusion, the statistical mechanics treatment of complex networks shown in this paper is able to put in a similar context, but in opposite limits, the emergence of scale-free networks and finite-scale networks. 
 Scale-free degree distribution correspond to higher energy states of the network respect to  finite-scale networks. Especially homogeneous random graphs obtained for $z=0$  have minimal  energy. Heterogenous scale-free networks correspond to  higher energy states. We have shown that this implies that uncorrelated  simple scale-free networks live in a  space of  allowed  networks relatively small. We believe that correlations present in various complex networks describing many technological, biological and social systems can only further reduce this space. Consequently the evolutionary rule by which this degree distribution is selected reveal a tendency of minimizing this space.

\end{document}